\documentclass[lineno]{jfm}

\usepackage{graphicx}
\usepackage{newtxtext}
\usepackage{newtxmath}
\usepackage{natbib}
\usepackage{amsmath}
\usepackage{hyperref}
\usepackage{siunitx}
\def\pd{\partial}
\hypersetup{
    colorlinks = true,
    urlcolor   = blue,
    citecolor  = black,
}

\newcommand{\RomanNumeralCaps}[1]
\linenumbers

\def\pd{\partial}

\title{Inferring rheology from free-surface observations}
\author{Edward M. Hinton\aff{1}\corresp{\email{ehinton@unimelb.edu.au}}}

\affiliation{\aff{1}School of Mathematics and Statistics, The University of Melbourne, Parkville, VIC 3010, Australia}

\begin{document}
\maketitle

\begin{abstract}
We develop direct inversion methods for inferring the rheology of a fluid from observations of its shallow flow. First, the evolution equation for the free-surface flow of an inertia-less current with general constitutive law is derived. The relationship between the volume flux of fluid and the basal stress, $\tau_b$ is encapsulated by a single function $F(\tau_b)$, which depends only on the constitutive law. The inversion method consists of (i) determining the flux and basal stress from the free-surface evolution, (ii) comparing the flux to the basal stress to constrain $F$ and (iii) inferring the constitutive law from $F$. Examples are presented for both steady and transient free-surface flows demonstrating that a wide range of constitutive laws can be directly obtained. For flows in which the free-surface velocity is known, we derive a different method, which circumvents the need to calculate the flux.
\end{abstract}

\section{Introduction}
\label{sec:intro}
In many environmental and industrial settings, it is important to determine the rheology of a fluid from observations of its free surface \citep{sellier2016inverse}. A major example is the simple slump test used in the mining, concrete and food industries to characterize particular features such as the yield stress of the fluid. In other contexts, such as for lava flows and ice sheets, deploying a rheometer may not be practical and instead free-surface observations provide an alternative constraint on the rheology \citep{roussel2005fifty,balmforth2007viscoplastic,martin2015inverse}.
\par
Previous methods for inferring the rheology of a fluid from its free surface typically involve comparing observations to simulated predictions \citep{piau2005consistometers,sayag2013axisymmetric,sellier2016inverse,al2019identification}. Simulated predictions are obtained from a forward model, which calculates the free-surface shape for given fluid properties. A particular form of the constitutive equation that has one or more free parameters, such as the Herschel-Bulkley law, is assumed. The free-surface evolution is calculated from this model over a sample of the parameter space and the mismatch between the simulated predictions and the observed data is minimised to constrain the rheological parameters.
\par
These `best-fit' methods rely upon a particular pre-supposed form of the constitutive law, which may be inaccurate or even unphysical at some values of the stress \citep{matsuhisa1965analytical,barnes1985yield,myers2005application}. They also require a large amount of forward computation. Finally, since the entire form of the constitutive law is determined by a global comparison to the observed flow, the parameters may be non-unique. The latter two of these concerns are exacerbated as the number of free parameters is increased.
\par
In this paper, direct inversion methods to infer general constitutive laws from free-surface data are described for inertia-less steady and transient flows. 
Our direct inversion methods provide conceptual insight into how predictions of the stress at the base of the fluid can be obtained from the free-surface elevation.
The constitutive law is reconstructed by comparing these stresses to the observed flux or velocity of the flow. In contrast to the indirect, global `best-fit' methods, the rheology is constrained in a pointwise and direct fashion. This reduces the computation required and enables a wide range of constitutive laws to be inferred.
\par
The first step in developing direct inversion methods is to derive a model for the flow thickness of a fluid with a given general constitutive law. We restrict our attention to inelastic, isothermal fluids with time-independent rheology (known as `generalized Newtonian fluids'). 
For two-dimensional flows, \citet{pritchard2015shallow} presented a governing equation for the flow thickness that is applicable to such rheologies. They exploited the fact that in shallow flows, the stress distribution is a known linear function of the vertical coordinate. The flux is written in terms of an integral over the stress that depends only on the constitutive law. In \S \ref{sec:forward}, we derive a forward model for a generalised Newtonian fluid in three dimensions by observing that the velocity and pressure gradient of shallow flows are parallel and their direction is independent of vertical position. This model suggests direct inversion methods for steady and transient flows (\S \ref{sec:invsteady} and \S \ref{sec:transient}). Concluding remarks are made in \S \ref{sec:conc}.

\section{Forward model} \label{sec:forward}
In this section, we develop a three-dimensional forward model for the flow of a generalised Newtonian fluid. The two-dimensional version was presented by \citet{pritchard2015shallow}. We analyse the flow of a generalized Newtonian fluid on topography with elevation given by $z=E(x,y)$. The fluid has thickness $h(x,y,t)$ and the free surface is at $z=E(x,y) + h(x,y,t)$. The effects of surface tension and inertia are neglected. The flow is assumed to be shallow; the thickness scale is much smaller than the lengthscales in the $x$ and $y$ direction. The vertical velocity, $w$, is negligible relative to the velocities ($u$, $v$) in the $x$ and $y$ directions. Using this lubrication approximation, the strain rate simplifies to
\begin{equation}
\dot{\gamma} =\Bigg\rvert \frac{\pd \boldsymbol{u}}{\pd z} \Bigg\rvert =\sqrt{(\pd u/\pd z)^2 + (\pd v/\pd z)^2},
\end{equation}
where $\boldsymbol{u} = (u,v)$.
As a constitutive law, we use a generalised Newtonian model with 
\begin{equation}
\dot{\gamma}= \phi(\tau),
\end{equation}
for a prescribed function $\phi(\tau)$, where $\tau=|\boldsymbol{\tau}|$ is the magnitude of the shear stress. This generalised form of the constitutive law is applicable to a wide range of inelastic fluids. Well-known constitutive laws such as the Newtonian, Bingham and Herschel-Bulkley models are special cases of this representation. 
\par
Under the shallow flow assumption, the pressure $p$ within the fluid is hydrostatic,
\begin{equation} \label{eq:PG}
p=\rho g (h+E -z), 
\end{equation}
where $\rho$ is the fluid density. The momentum equation in the $x$ and $y$ directions is given by
\begin{equation}\label{eq:PGtau}
\frac{\pd \boldsymbol{\tau}}{\pd z} = - \boldsymbol{G},  \qquad \text{where} \qquad \boldsymbol{G} = -\nabla p = -\rho g \nabla (h+E).
\end{equation}
The vector $\boldsymbol{G}$ is the (negative) hydrostatic pressure gradient and $\nabla = (\pd/\pd x, \pd/\pd y)$. Upon integrating with respect to $z$ we obtain
\begin{equation} \label{eq:momint}
\boldsymbol{\tau} = (h+E-z)\boldsymbol{G},
\end{equation}
since there is zero stress at the free surface. The stress is a linear function of the vertical position within the flowing layer. The volume flux is rearranged as follows \citep{schwartz2002flow,pritchard2015shallow}
\begin{equation} \label{eq:Qinit}
\boldsymbol{Q} = \int_{E}^{E+h} \boldsymbol{u} \, \mathrm{d} z = \int_{E}^{E+h} (h+E-z) \frac{\pd \boldsymbol{u}}{\pd z} \, \mathrm{d} z,
\end{equation}
where we have used no slip at the base, $z=E$. Note that the vectors, $\boldsymbol{G}$, $\boldsymbol{\tau}$, $\boldsymbol{u}$ and $\boldsymbol{Q}$ are parallel with the same direction (in the $(x,y)$ plane) as the flow. Since the pressure gradient, $\boldsymbol{G}$ is independent of $z$, the direction of $\boldsymbol{\tau}$ and $\boldsymbol{u}$ is also independent of $z$ (as is the direction of $\pd \boldsymbol{u}/\pd z$ and $\pd \boldsymbol{\tau}/\pd z$). We take  the magnitude of both sides of \eqref{eq:Qinit} and these observations allow us to take the magnitude through the integral (the direction of the integrand is independent of $z$). Next, we substitute for $h+E-z$ from \eqref{eq:momint} in \eqref{eq:Qinit}. Also, equation \eqref{eq:momint} furnishes the relation $- |\boldsymbol{G}| \mathrm{d} z= \mathrm{d} \tau$ and the magnitude of the flux becomes an integral over the stress within the layer,
\begin{equation} \label{eq:fluxmag1}
|\boldsymbol{Q}| = |\boldsymbol{G}|^{-2} \int_0^{|\boldsymbol{G}|h} \tau \phi(\tau) \, \mathrm{d} \tau.
\end{equation}
The quantity, $|\boldsymbol{G}|h$ is the magnitude of the stress at $z=E(x,y)$ (the basal stress). In this formulation, the flux has been written in terms of an integral of the stress, $\tau$, and the shear rate, $\phi(\tau)$.
\par
Since the flux and pressure gradient are parallel (in the $(x,y)$ plane), we write
\begin{equation} \label{eq:fluxdef}
\boldsymbol{Q} = \frac{F(|\boldsymbol{G}| h)}{|\boldsymbol{G}|^3} \boldsymbol{G}, \qquad \text{where} \quad F(\tau_0) = \int_0^{\tau_0} \tau \phi(\tau) \, \mathrm{d} \tau.
\end{equation}
A similar equation was derived by \citet{schwartz2002flow} with the integrand written in terms of a stress-dependent viscosity.
The function $F(\tau_0)$ depends only on the rheology of the fluid. It accounts for how the volume flux varies with the basal stress. \citet{schwartz2002flow} describe this variation as the `fluidity' of the flow.
Mass continuity furnishes the governing equation for the flow thickness,
\begin{equation} \label{eq:forwardtransient}
\frac{\pd h}{\pd t} + \nabla \cdot \Big[ |\boldsymbol{G}|^{-3} F(|\boldsymbol{G}| h) \boldsymbol{G} \Big] = 0,
\end{equation}
which completes the forward model for $h(x,y,t)$ since the pressure gradient $\boldsymbol{G}$ depends linearly on the gradient of the flow thickness. In the case that $\pd h/\pd y=\pd E/\pd y=0$, the two-dimensional governing equation of \citet{pritchard2015shallow} is recovered. Since we are interested in general constitutive laws, there is no intrinsic scaling for the stress and strain rate so we keep the problem dimensional throughout this paper. 
For steady flows ($\pd h/\pd t=0$), the forward problem is solved with boundary conditions for $h(x,y)$ at the edges of a particular domain. The system is integrated numerically and details of the method are given in Appendix \ref{app:numerics}. The solution for steady flow over a topographical feature is shown in figure \ref{fig:oneDtopog}a,b for a specific constitutive law, given in figure \ref{fig:oneDtopog}d. Corresponding plots for the steady flow around a cylinder on an inclined plane for a different constitutive law are shown in figure \ref{fig:cylinder}.

\section{Inversion methods for steady flows} \label{sec:invsteady}
The steady inverse problem involves inferring the rheology of the fluid (i.e. the function, $\phi(\tau)$), given the elevation of the topography, the fluid free surface and its density. To obtain the constitutive law, the function $F(\tau)$ is reconstructed. We rewrite \eqref{eq:fluxdef} as
\begin{equation} \label{eq:Fdefinv}
F(|\boldsymbol{G}| h) = |\boldsymbol{Q}| |\boldsymbol{G}|^2.
\end{equation} 
The quantities $|\boldsymbol{Q}|$, $|\boldsymbol{G}|$ and $h$ (the flux, pressure gradient and flow thickness) are determined at various points in the flow, each of which provides a prediction of $F(\tau)$ at $\tau =|\boldsymbol{G}| h$. 
We calculate these three quantities as follows. First, we measure the flow thickness, $h$, from the difference between the free-surface elevation and the (known) topographic elevation. The pressure gradient, $|\boldsymbol{G}|$, is calculated from the gradients of the free surface using \eqref{eq:PGtau}b. Finally, to determine the flux, $|\boldsymbol{Q}|$, the steady version of the governing equation \eqref{eq:forwardtransient} is rewritten as
\begin{equation}
\nabla \cdot \Big(|\boldsymbol{Q}|\boldsymbol{\hat{G}} \Big) =0, \qquad \text{where} \quad \boldsymbol{\hat{G}}=\boldsymbol{G}/|\boldsymbol{G}|,
\end{equation}
and $\boldsymbol{\hat{G}}$ is the unit vector in the $(x,y)$ plane in the direction of the flow. This equation is solved for $|\boldsymbol{Q}|$ rather than for $h$ as in the forward problem. The method of characteristics furnishes
\begin{equation} \label{eq:chars}
\frac{\mathrm{d} \boldsymbol{x}}{\mathrm{d} s} = \hat{\boldsymbol{G}}, \qquad \frac{\mathrm{d} |\boldsymbol{Q}|}{\mathrm{d} s} = - |\boldsymbol{Q}| \nabla \cdot \hat{\boldsymbol{G}},
\end{equation}
where $\boldsymbol{x}=(x,y)$, and $s$ parameterises the characteristic curves in the $(x,y)$ plane, which are the streamlines of the flow. A similar method was used by \citet{sellier2010beating} for reconstructing bottom topography from the free-surface flow of thin films of Newtonian fluid \citep[see also][]{heining2012inverse}. The direction of the streamlines is given by the (known) direction of $\boldsymbol{G}$. Then by integrating \eqref{eq:chars}, the magnitude of the flux $|\boldsymbol{Q}|$ is obtained along the streamlines.
\par
The integration along characteristics requires the value of $|\boldsymbol{Q}|$ at some upstream location. If we only have the steady free-surface elevation and we have no data concerning the timescale of the flow (such as the flux or free-surface velocity), then the inverse problem is incomplete. In its simplest form, this incompleteness arises in the problem of determining the viscosity of a Newtonian flow from its steady free-surface elevation; one must know something of the timescale to constrain the shear rate. In general, without knowing the flow timescale, the constitutive law can be determined up to a multiplicative constant (essentially a reference viscosity).
\par
In some cases, the flux far upstream may be known or could at least be estimated. In the case where the far upstream flux is not known, but there is some other data concerning the timescale of the flow, we make an arbitrary choice of $|\boldsymbol{Q}|=1 \si{m^2/s}$ far upstream, which will subsequently be corrected. We then determine $|\boldsymbol{Q}|$ along the streamlines, from which we calculate $F$ and $\phi$ as described below. Owing to the linearity of \eqref{eq:chars}b and the linear relationship between $|\boldsymbol{Q}|$,  $\phi$ and $F$, the incorrect upstream flux merely leads to a constant multiplicative error in these quantities. This multiplicative constant may be constrained by comparison with observed data of the flux or free-surface velocity from at least one point along the streamline (e.g. by using the formulation in \S \ref{subsec:freesurfvel}).

\begin{figure}
\includegraphics{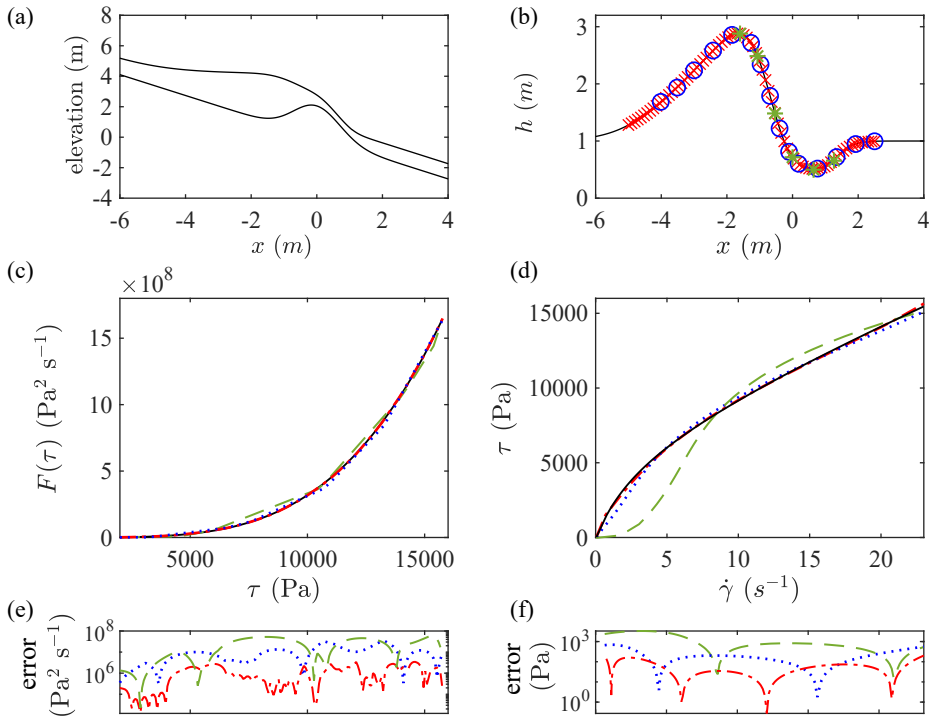}
\caption{Forward and inversion method for steady flow over topography. (a) Topographic and free-surface elevation. (b) Flow thickness and locations of the points used for the inversion method. (c) Predictions of $F(\tau)$ (dotted lines) and its true form (continuous line). (d) Reconstructed constitutive law for the three sets of points in (b) is compared to its true shape (\ref{eq:constittopog}, solid line). (e,f) Absolute error between the inferred and true quantities in panels (c) and (d).}
\label{fig:oneDtopog}
\end{figure}

We use the data (of $|\boldsymbol{Q}|$, $|\boldsymbol{G}|$ and $h$) along the streamlines to infer the relationship between $|\boldsymbol{G}| h$ and $|\boldsymbol{Q}| |\boldsymbol{G}|^2$, which provides an approximation for $F(\tau)$ using \eqref{eq:Fdefinv}. For a given stress, $\tau'$, we determine $F(\tau')$ from observations of the flow at a point where the basal stress equals $\tau'$. With a prediction for $F(\tau)$ in hand, the constitutive relation is reconstructed via
\begin{equation} \label{eq:phiFderiv}
\dot{\gamma} = \phi(\tau) = \frac{1}{\tau} \frac{\mathrm{d} F}{\mathrm{d} \tau}.
\end{equation}
This requires a prediction of the derivative of $F$ from scattered data; we either fit a polynomial or a piecewise linear function (e.g. figure \ref{fig:oneDtopog}c). We can only predict the function $F(\tau)$, and hence the constitutive law $\phi(\tau)$, over the range of values that the basal shear stress, $|\boldsymbol{G}|h$, takes within the flow. The more the steady flow is diverted by the topography or obstruction, the greater the range of $\tau$ over which the constitutive law can be inferred. Further details of the method are provided in the following subsections by studying two examples; flow over topography (\S \ref{subsec:steadytopog}) and flow around a cylinder (\S \ref{subsec:steadycyl}). In each case, we assume that the upstream flux is known.

\subsection{Rheology from steady flow over topography} \label{subsec:steadytopog}
We deploy the inversion method to show how the constitutive law can be inferred from the shape of the free surface that arises in steady flow over a topographic feature. The topography consists of a slope at an angle $\beta=0.6$ to the horizontal with a Gaussian mound added at the origin (shown in figure \ref{fig:oneDtopog}a). The feature extends to infinity in the positive and negative $y$ directions and hence the problem is two-dimensional. The flow is in the $x$ direction and the volume flux, $|\boldsymbol{Q}|$, is constant along the $x$ axis. 
The forward problem for the flow thickness is solved numerically as described in Appendix \ref{app:numerics} with the following constitutive relation (black line in figure \ref{fig:oneDtopog}d)
\begin{equation} \label{eq:constittopog}
\dot{\gamma}=\phi(\tau)= \Bigg(1+ \frac{M-1}{\sqrt{1+a^2 \tau^2}}\Bigg) \frac{\tau}{\mu},
\end{equation}
where $M=0.2$, $\mu=400\, \si{\pascal.s}$ and $a=1.1\, \times 10^{-4}\, \si{\pascal^{-1}}$.
The flux far upstream is $1.67 \,\si{m^2/s}$ per unit width and the fluid density is $\rho = 1200\, \si{kg.m^{-3}}$. The flow thickness far upstream is thus 1 metre. The calculated free-surface elevation is shown relative to the topography in figure \ref{fig:oneDtopog}a and the flow thickness is plotted in figure \ref{fig:oneDtopog}b. The flow deepens (and slows) upstream of the mound, where the basal stress is small \citep[c.f.][]{hinton2019interaction}. The flow is thinner and faster on the downstream side of the mound and the basal stress is greater.
\par
The inversion method is carried out for three sets of observations of the synthetic forward prediction. The first inversion uses the free-surface gradient, free-surface elevation and topographic elevation at 80 points along the $x$ axis, shown by red crosses in figure \ref{fig:oneDtopog}b. The function $F$ is predicted from this data (see equation \ref{eq:Fdefinv}). At each of the 80 points, we determine the flow thickness, $h$ and the pressure gradient, $|\boldsymbol{G}|$ from the gradient of the free surface (calculated via a finite difference approximation using the 80 points). The flux is a known constant along the $x$ axis. From these measurements, we calculate the basal shear stress, $|\boldsymbol{G}|h$ and $|\boldsymbol{G}|^2 |\boldsymbol{Q}|$ at the 80 points. This data is sorted from the smallest to the largest basal stress and plotted using linear interpolation. This prediction is shown as a red dashed line in figure \ref{fig:oneDtopog}c, which shows close agreement with the true $F(\tau)$ (continuous black line). The prediction of $F$ furnishes an estimate of $\phi(\tau)$ via \eqref{eq:phiFderiv}. This is plotted as a red dashed line in figure \ref{fig:oneDtopog}d and shows excellent agreement with the true constitutive law (continuous black line). The same method is then carried out with 15 sample points (blue circles and blue dotted lines) and 6 sample points (green stars and green dashed lines) demonstrating that good predictions of the constitutive law may be obtained with fifteen observations focused in the region where the flow thickness varies significantly. Figure \ref{fig:oneDtopog} illustrates that when there are fewer data points (e.g. the green curves), small errors in $F$ can lead to more significant errors in the constitutive law because the derivative of $F$ is not well-approximated. The error is larger at smaller values of $\tau$ owing to the $1/\tau$ pre-factor in \eqref{eq:phiFderiv}.
\par
Finally, it should be noted that this method requires the topographic elevation. In the context of ice sheets, this is frequently obtained via either airborne or ground-based radio-echo sounding. For lava flows, the pre-eruption topography is often known. The other examples in this paper do not require such detailed topographic data (e.g. slump tests on a horizontal plane).

\subsection{Rheology from steady flow around cylinders} \label{subsec:steadycyl}
\begin{figure}
\includegraphics{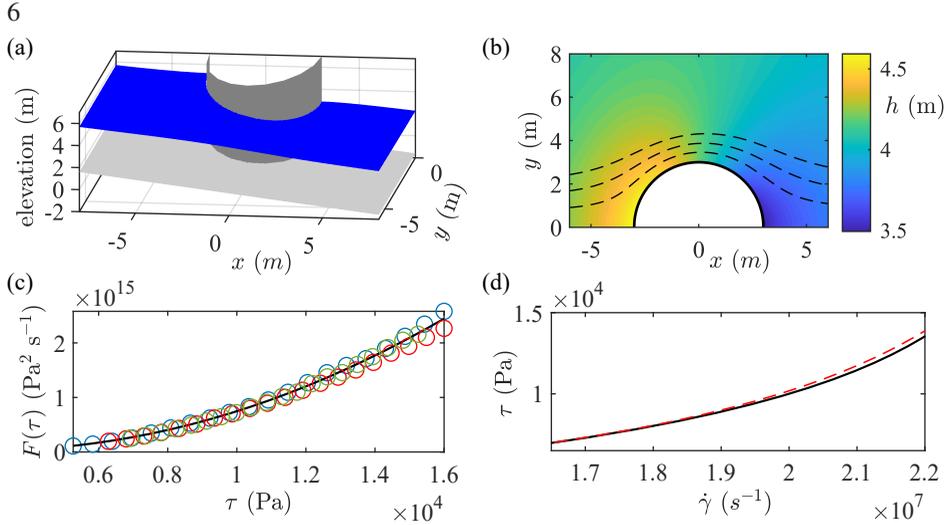}
\caption{Steady flow around a cylinder. (a) Free-surface elevation (blue) relative to the cylinder and plane. (b) Flow thickness and three streamlines (dashed lines). (c) Predictions of $F(\tau)$ from the streamlines (circles) and its true form (continuous line). (d) Prediction of the constitutive law (red dashed line) and its true form (\ref{eq:constitcyl}, continuous line).}
\label{fig:cylinder}
\end{figure}
In this subsection, we infer the rheology of a fluid from observations of its steady flow around a circular cylinder on an inclined plane with the axis of the cylinder in the $z$ direction (see figure \ref{fig:cylinder}). The fluid has the following constitutive law (black line in figure \ref{fig:cylinder}d),
\begin{equation} \label{eq:constitcyl}
\dot{\gamma}=\phi(\tau) = \frac{\tau^2/\mu_2}{1+a^2\tau^2},
\end{equation}
with $\mu_2 =10^7\, \si{\pascal^2.s}$ and $a=2\,\times 10^{-4}\,  \si{\pascal^{-1}}$. The plane is inclined with gradient $\tan \beta=0.2$. Far upstream of the cylinder, the steady thickness is $H=4$ metres and the flux per unit width is $9.74 \, \si{m^2/s}$. The cylinder radius is 3 metres and the fluid has density $\rho =1200\, \si{kg.m^{-3}}$. The forward problem is solved numerically as described in Appendix \ref{app:numerics} with no-flux boundary condition on the cylinder. The flow is symmetric about the centreline (which is the $x$ axis). Figure \ref{fig:cylinder}a shows the free-surface elevation and figure \ref{fig:cylinder}b shows the flow thickness, which increases by approximately 15\% upstream of the cylinder and decreases by a similar amount downstream \citep[c.f.][]{hinton2020viscous}.
\par
The solution for the free-surface elevation is used as the input for the inversion method. The free-surface elevation and its gradient are known on a discrete grid with $10\,\si{cm}$ spacing in the $x$ and $y$ directions. The pressure gradient can be obtained from the free-surface gradient and the flux is calculated along three streamlines using \eqref{eq:chars} (dashed lines in figure \ref{fig:cylinder}b). The prediction for $F(\tau)$ from each streamline is shown as a different set of coloured circles in figure \ref{fig:cylinder}c. There is good agreement with the true form (black line). A fourth order polynomial is fitted to the three sets of data from which we obtain an inference of the constitutive law via \eqref{eq:phiFderiv}. Figure \ref{eq:phiFderiv}d compares the predicted constitutive law (red dashed line) to its true form (black line). The range of strain rates over which the constitutive law is inferred is fairly small because the cylinder is relatively thin leading to only small variations in the basal stress. Observations of flow around a wider cylinder would provide predictions of $\dot{\gamma}=\phi(\tau)$ over a larger range.

\section{Inversion methods for transient flows} \label{sec:transient}
For the forward transient problem, the governing equation \eqref{eq:forwardtransient} is solved with the relevant initial and boundary conditions and a source term may be included on the right-hand side in the case of a supplied input flux. Details of the transient forward numerical method are given in Appendix \ref{app:numerics}.
\par
In the following two subsections (\S \ref{subsec:flux} and \S \ref{subsec:freesurfvel}), two inversion methods for obtaining the rheology from transient free-surface flow are described and examples are given in each case. The method of \S \ref{subsec:flux} is relevant to two-dimensional and axisymmetric flows. We determine the flux at a given location and compare this to the pressure gradient to predict $F(\tau)$ in an analogous fashion to the steady method. \S \ref{subsec:freesurfvel} introduces a different approach, which utilises the free-surface velocity of the flow.

\subsection{Flux-based method} \label{subsec:flux}
\begin{figure}
\includegraphics{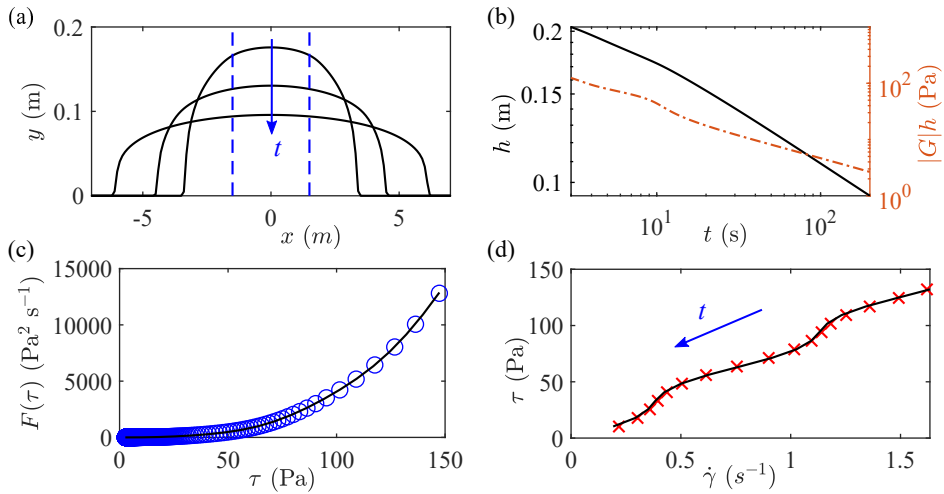}
\caption{(a) Slump shape at 12.5, 50 and 200 seconds (black lines) and the edges of a control rectangle, $x=\pm 1.5$ metres (blue dashed lines). (b) Flow thickness (black line, left axis) and basal stress (dashed line, right axis) at the edge location. (c) Prediction of $F$ (circles) versus true $F$ (solid line). (d) Inferred constitutive law (red crosses) and the true constitutive law, \eqref{eq:rheolsine} (solid line).}
\label{fig:sineslump}
\end{figure}
For a two-dimensional flow, the flux across any given station, $x=L$, can be obtained from the rate of change of the volume in $x<L$, which in turn can be calculated from the evolution of the free surface. The flux at $x=L$ can be written as
\begin{equation} \label{eq:Qcalc1}
Q(x=L,t)= q_s - \frac{\pd}{\pd t} \Bigg( \int_{-\infty}^L h \, \mathrm{d} x \Bigg),
\end{equation}
where $q_s$ is any source flux in $-\infty<x<L$ if it exists. Given the free-surface elevation at various times, the flux, the flow thickness and the pressure gradient can be obtained at $x=L$ at those times. The function $F(\tau)$ is reconstructed by using equation \eqref{eq:Fdefinv}. Unlike the steady method, additional data about the upstream flux or free-surface velocity is not required to fully constrain $F(\tau)$ (and the constitutive law) because the transient evolution of the free surface provides sufficient data to determine the timescale of the flow.
\par
As an example, we consider a two-dimensional slump of material, with volume per unit length of $1\, \si{m^2}$, released as a square block on a horizontal plane. Slumps of Herschel-Bulkley fluids on a horizontal plane were described in detail by \citet{balmforth2000visco}. We use the following constitutive law
\begin{equation} \label{eq:rheolsine}
\dot{\gamma}=\phi(\tau)= \frac{\tau}{\mu_0} + \alpha \sin(\tau/\tau^{*}),
\end{equation}
with $\tau^{*}= 10\, \si{\pascal}$, $\alpha=0.08\, \si{s^{-1}}$ and $\mu_0=83\, \si{ \pascal.s}$. Figure \ref{fig:sineslump}a shows the free-surface evolution, calculated numerically using the forward model \eqref{eq:forwardtransient}. The flow is symmetric about $x=0$. 
\par
For the inversion method, we use the free-surface elevation data at 10 $\si{cm}$ intervals along the $x$ axis and half second intervals in time. Equation \eqref{eq:Qcalc1} is used to estimate the flux at $x=\pm 1.5$ metres (blue dashed line in figure \ref{fig:sineslump}a). Figure \ref{fig:sineslump}b shows the flow thickness and the basal stress, $|G|h$, calculated from the gradient of the free surface, at $x=L=1.5$ metres as a function of time. The undulations arise from the nonlinearities in the constitutive law. The basal stress, flow thickness and flux provide a prediction for $F(\tau)$, which is plotted with blue circles in figure \ref{fig:sineslump}c and compared to the true shape of $F$ (black line). The derivative of $F$ is obtained via linear interpolation and this furnishes a prediction of the constitutive law (red crosses in figure \ref{fig:sineslump}d). Provided the flow thickness in $-L<x<L$ and its gradient at $x=L$ can be obtained to sufficient accuracy at multiple times, the inversion method is able to capture the detailed nonlinearities in the constitutive law. The constitutive law is determined for larger stresses at earlier times and smaller stresses at late times as the fluid slumps (see figure \ref{fig:sineslump}b).

\begin{figure}
\includegraphics{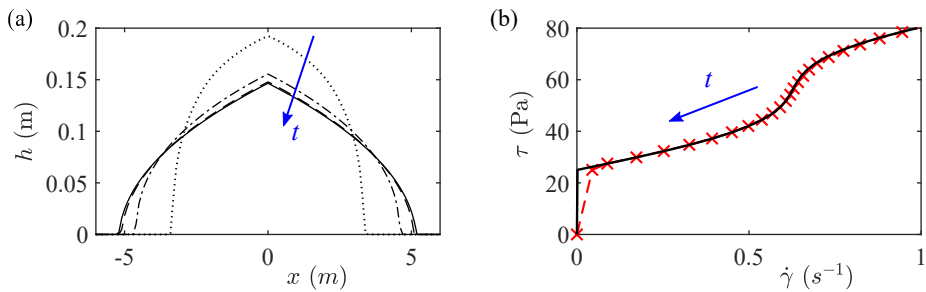}
\caption{(a) Free-surface shape for a yield-stress fluid at $10$, $10^2$, $10^3$ and $10^4$ seconds. (b) The constitutive law (black line) and inference from the inversion method (red crosses).}
\label{fig:sineyield}
\end{figure}

The inversion method can also be applied to a material with a yield stress. We repeat the forward solution from above but the fluid now has a yield stress of $25 \, \si{\pascal}$. 
The free surface is shown at four times in figure \ref{fig:sineyield}a. At late times, the stationary slumped shape is approached. The inversion method is applied to predict the constitutive law and we include the additional constraint that $\phi(0)=0$ (see figure \ref{fig:sineyield}b). The basal stress at $x=L$ reduces in time and slowly converges to the yield stress (algebraically with respect to time \citep[c.f][]{hogg2009slumps}). Hence, to capture the constitutive law close to the yield stress, the slumping fluid must be observed for a long time. The yield stress is accurately captured at $t=10^4$ seconds (see figure \ref{fig:sineyield}b).
In previous research, the final slump shape has been used to obtain the yield stress and our method can reproduce those results as well as capturing the constitutive law at other strain rates from the transient evolution \citep{roussel2005fifty,balmforth2007viscoplastic}.
Finally, it should be noted that the method of this section is easily adapted to axisymmetric flows.

\subsection{Utilising the free-surface velocity} \label{subsec:freesurfvel}
\begin{figure}
\includegraphics{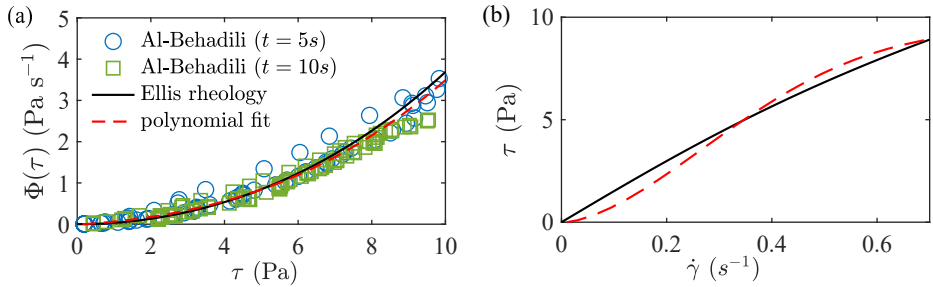}
\caption{Inferring the constitutive law from the free-surface velocity data of \citet{al2019identification}. (a) $\Phi(\tau)$ predicted using our inversion method applied to their data (circles and squares) with a polynomial best fit (red dashed line) and the correct shape (black continuous line). (b) Inferred (red dashed line) and true (black line) constitutive law.}
\label{fig:sellierdata}
\end{figure}

For unsteady three-dimensional flows, a different inversion method is required. In this section, we show how the free-surface velocity can be written in terms of the rheology of the fluid. This formulation provides a basis for an inversion method that uses the free-surface elevation and velocity to determine the fluid rheology. Such a method is ideal for exploiting experimental Particle Image Velocimetry (PIV) data.
\par
We write the magnitude of the free-surface velocity in the form
\begin{equation}
| \boldsymbol{u}(z=E+h)| = \int_E^{E+h} \bigg\rvert \frac{\pd \boldsymbol{u}}{\pd z} \bigg\rvert \, \mathrm{d} z,
\end{equation}
where the magnitude is taken through the integral because the direction of the integrand is independent of $z$ (see \S \ref{sec:forward}).
We apply the change of variables $-|\boldsymbol{G}| \mathrm{d} z = \mathrm{d} \tau$ to obtain the following relationship between the free-surface velocity and the fluid rheology,
\begin{equation}
|\boldsymbol{u}(z=E+h)|=|\boldsymbol{G}|^{-1} \Phi(|\boldsymbol{G}| h), \qquad \Phi(\tau_0)= \int_0^{\tau_0} \phi(\tau)\, \mathrm{d} \tau.
\end{equation}
A similar formulation can be applied to obtain the velocity at each height within the layer.
The inversion method is similar to that of previous sections. We constrain $\Phi(\tau)$ by comparing the observed free-surface velocity of the flow,  $|\boldsymbol{u}(z=E+h)|$, to the pressure gradient, $|\boldsymbol{G}|$, and basal stress, $|\boldsymbol{G}|h$ at various locations and times. The constitutive law is simply recovered from $\phi(\tau)=\mathrm{d} \Phi/\mathrm{d} \tau$.
\par
By way of an example, we apply this inverse approach to the data of \citet{al2019identification}. Their figures 5 and 6 show the flow thickness and free-surface velocity at three times for a dam break of an Ellis fluid calculated from a full Navier-Stokes simulation. 
First, we calculate the basal stress and pressure gradient at two times and thirty-five spatial points from the flow thickness and its gradient. The data for the basal stress, $|\boldsymbol{G}| h$, is plotted against the observed free-surface velocity multiplied by the magnitude of the pressure gradient, $|\boldsymbol{G}|$, at each point to obtain a prediction for $\Phi(\tau)$, shown as scattered squares and circles in figure \ref{fig:sellierdata}a. The true form of $\Phi(\tau)$ is shown as a solid black line. A fourth order polynomial (red dashed line in figure \ref{fig:sellierdata}a) is fitted to the scattered data with the condition that its derivative vanishes at the origin, which ensures that $\phi(0)=0$. This polynomial is differentiated to obtain a prediction for the constitutive law (red dashed line in figure \ref{fig:sellierdata}b), which shows reasonable agreement with the constitutive law used in the simulations of \citet{al2019identification}. The simulations included relatively small amounts of inertia and surface tension, which are not accounted for in our inversion method, although the latter could be included in the pressure gradient. Despite this, our simple direct inversion method reproduces the rheology to within a 15\% error over most shear rates.

\section{Conclusion} \label{sec:conc}
This contribution has presented direct inversion methods for determining the constitutive law for a fluid from observations of its free surface. For steady flows, the flux along streamlines is compared to the pressure gradient and flow thickness to constrain a function, $F$, that encapsulates the rheology. We have also presented a technique for inferring the constitutive law from the free-surface velocity. The methods could be extended to incorporate an undetermined sliding law at the bed as is relevant to the migration of ice sheets. In other settings, the flow may be compressible or inertia may be important. In these cases, the relationship between $|Q|$, $F$ and $\phi$ may be nonlinear, which would lead to a more complicated inversion method. Further work could also study optimum methods for taking the derivative of $F$ from noisy data.
\par
\textbf{Declaration of Interests}: The author reports no conflict of interest.

\section*{Acknowledgements}
The author is grateful to the School of Mathematics and Statistics at the University of Melbourne for the award of a Harcourt-Doig research fellowship.

\appendix
\section{Numerical methods} \label{app:numerics}
In this appendix, the numerical methods for the forward problems are described. The flow thickness for two-dimensional steady flow over topography (figure \ref{fig:oneDtopog}) is calculated numerically using equation \eqref{eq:fluxdef}. The flux is set to a constant value, which furnishes a first order ordinary differential equation for $h$. This is integrated numerically in the upstream direction using a fixed value of the flow thickness far downstream; a numerical instability arises from integration in the downstream direction \citep[see][]{hinton2019interaction}. The steady three-dimensional flow around a cylinder (figure \ref{fig:cylinder}) is solved numerically using MATLAB's Partial Differential Equation (PDE) Toolbox\textsuperscript{TM}. The method is given in section III of \citet{hinton2020viscous} with the function inside the divergence operator adjusted to account for the general rheology. 
For this flow, $h>0$ and $|\boldsymbol{G}|>0$ and no regularisation is required.
\par
For transient flows (figures \ref{fig:sineslump} and \ref{fig:sineyield}), the forward problem \eqref{eq:forwardtransient} is solved using the toolbox's transient solver. In the case that the fluid has a yield stress, a regularisation of the constitutive law is required. If the constitutive law without a yield stress is $\dot{\gamma}=\phi(\tau)$ then for a fluid with yield stress, $\tau_Y$, we use the following regularisation
\begin{equation}
\dot{\gamma}= \phi \Big\{\Big[\tau-\tau_Y + \sqrt{(\tau-\tau_Y)^2 + \epsilon^2 \tau_Y^2}\Big]/2 \Big\},
\end{equation}
where $\epsilon=10^{-5}$. We also replace $|\boldsymbol{G}|$ with $\sqrt{|\boldsymbol{G}|^2+(\delta \rho g)^2}$ where $\delta =10^{-8}$. Checks confirm that volume is conserved. The inversion methods provide an additional check on the forward computations.

\bibliographystyle{jfm}
\bibliography{jfm}


\end{document}